\documentclass[conference]{IEEEtran}
\IEEEoverridecommandlockouts
\usepackage{cite}
\usepackage{amsmath,amssymb,amsfonts}
\usepackage{algorithmic}
\usepackage{graphicx}
\usepackage{textcomp}
\usepackage{xcolor}
\def\BibTeX{{\rm B\kern-.05em{\sc i\kern-.025em b}\kern-.08em
    T\kern-.1667em\lower.7ex\hbox{E}\kern-.125emX}}

\begin{document}

\title{A Stochastic Geo-spatiotemporal Bipartite Network to Optimize GCOOS Sensor Placement Strategies
}

\author{\IEEEauthorblockN{Ted Holmberg}
\IEEEauthorblockA{\textit{Department of Computer Science} \\
\textit{University of New Orleans}\\
New Orleans, LA, USA \\
eholmber@uno.edu}
\and
\IEEEauthorblockN{Elias Ioup}
\IEEEauthorblockA{\textit{Center for Geospatial Sciences} \\
\textit{Naval Research Laboratory,}\\
\textit{Stennis Space Center}\\
Mississippi, USA. \\
elias.ioup@nrlssc.navy.mil}
\and
\IEEEauthorblockN{Mahdi Abdelguerfi}
\IEEEauthorblockA{\textit{Cannizaro-Livingston Gulf States} \\
\textit{Center for Environmental Informatics}\\
New Orleans, LA, USA\\
chairman@cs.uno.edu}
}

\IEEEoverridecommandlockouts
\IEEEpubid{\footnotesize\begin{minipage}{\textwidth}\ \\[12pt]
\hrule width 0.49\columnwidth height 0.5pt \vspace{10pt}
Accepted version. \copyright 2022 IEEE. Personal use of this material is permitted. \\
Permission from IEEE must be obtained for all other uses.\\
DOI 10.1109/BigData55660.2022.10020928
\end{minipage}}

\maketitle

\begin{abstract}
This paper proposes two new measures applicable in a spatial bipartite network model: coverage and coverage robustness. The bipartite network must consist of observer nodes, observable nodes, and edges that connect observer nodes to observable nodes. The coverage and coverage robustness scores evaluate the effectiveness of the observer node placements. This measure is beneficial for stochastic data as it may be coupled with Monte Carlo simulations to identify optimal placements for new observer nodes. In this paper, we construct a Geo-SpatioTemporal Bipartite Network (GSTBN) within the stochastic and dynamical environment of the Gulf of Mexico. This GSTBN consists of GCOOS sensor nodes and HYCOM Region of Interest (RoI) event nodes. The goal is to identify optimal placements to expand GCOOS to improve the forecasting outcomes by the HYCOM ocean prediction model. 
\end{abstract}


\section{Introduction}
The Gulf of Mexico (GoM) is environmentally and economically vital to the US. Its coastline extends across five U.S. states: Texas, Louisiana, Mississippi, Alabama, and Florida \cite{b3}. It hosts multiple major ports and transportation waterways which provide the US with many critical resources: oil, gas, wind, waves, and seafood \cite{b2}. A diverse group of commercial, academic, federal, and local organizations jointly support operations in the GoM to observe, measure, and study the region. However, the GoM is vast, with over 17,000 miles of shoreline, and its basin encompasses 600,000 square miles.  Despite the concerted effort between organizations to build a shared sensor array, the current number of sensors only observes a sparse fraction of the GoM \cite{b1}. It is critical to supply these institutions with guidance on where optimal new sensor placements may go to best contribute to the sensor array. This problem becomes even  more challenging, considering the GoM continuously changes states.  Unlike land terrain, which remains relatively stable, water bodies are dynamic systems \cite{b5}. Network models and analysis provide key insights into where to place new GCOOS sensors.   

\section{Background}
Various ongoing initiatives are engaged in monitoring and reporting both historical and real-time states of the GoM. This paper focuses on two: The Gulf of Mexico Coastal Ocean Observing System (GCOOS) and HYbrid Coordinate Ocean Model (HYCOM). 

\subsection{GCOOS}
The GCOOS is the Gulf of Mexico regional component of the U.S. Integrated Ocean Observing System (IOOS). It is the only certified comprehensive data collection and dissemination center for coastal and ocean data in the Gulf. GCOOS collects data from 1,655 sensors located at 163 non-federal and 159 federal stations \cite{b1}.

\subsection{HYCOM}
HYCOM is a real-time three-dimensional grid mesh ocean model with 1/25° horizontal resolution that provides eddy-resolving hindcast, nowcast, and forecast as numerical states of the GoM.  HYCOM assimilates data from various sensors, such as satellites, buoys, ARGO floats, and autonomous underwater gliders (AUGs).  The forecast system is the Navy Coupled Ocean Data Assimilation (NCODA), a multivariate optimal interpolation scheme that assimilates surface observations. By combining these observations via data assimilation and using the dynamical interpolation skill of the model, a three-dimensional ocean state can be more accurately nowcast and forecast \cite{b4}.

\section{Objective}
This research aims to construct a GSTBN model using HYCOM and the observational sensor data from GCOOS. This GSTBN aims to identify regions of interest within the HYCOM model to recommend how best to utilize the sensor array of GCOOS and provide guidance on where to expand it.

\section{Motivation}
Localized regions of temporal variability within HYCOM hinder the accuracy of its nowcast/forecast. A region of temporal variability is where significant changes in a numerical property occur within the same coordinate between two consecutive temporal frames\cite{b6}. HYCOM produces nowcasts and forecasts by combining its real-time observations and prior historical data.   The forecasting error rate generally increases as the values between snapshots differ \cite{b6}. The best approach to mitigate such regions of temporal variability is to acquire new observations to feed into HYCOM \cite{b6}.  The next set of nowcasts and forecasts will then use the most up-to-date measures and ensure the error rate is as minimal as possible. By placing instruments into the regions of interest (RoI), GCOOS can get the data needed to maximize the accuracy rate in the HYCOM nowcasting and forecasting model.  However, the number of sensors is limited and proper planning should maximize their effectiveness in improving the nowcasting and forecasting model. 

\section{Approach}
By modeling a GSTBN composed of a set of observable nodes representing the RoI within the GoM and a set of observer nodes representing the GCOOS sensors. An RoI is identified by taking a set of temporal snapshots from HYCOM and computing the residuals over time, where the residual is the magnitude difference between snapshots. The nodes representing the GCOOS sensors have attributes consistent with that particular instrument, such as its operational status, geo-coordinates, current data readings, mobility speed, and institutional membership. The GSTBN establishes a link between the sensor nodes to all potential nearby RoI nodes. It facilitates the decision-making to assess which location to recommend planning for new installations for sensors, for relocation, or when to perform maintenance. Other decisions involving the sensor array might be when and where to grow the sensor network. Monte Carlo simulations identify optimal sensor placements by attempting to add a GCOOS node to the GSTBN randomly, and its effectiveness is then subsequently evaluated.

\section{Methods}
This is a stochastic problem, therefore there is no deterministic solution, and it is best to rely on random samplings to construct temporal graph representations by evaluating potential outcomes between GCOOS sensor placements and RoI positions. Graph analysis identifies and selects the optimal positions to maximize the GCOOS coverage and its coverage robustness.   

\subsection{Temporal Graph Representation}
GSTBN is a type of temporal graph. A temporal network is an ordered set of static graphs. The ordering is the static network's temporal occurrence or "snapshot" at a particular timestamp.
\begin{equation}
TG = \Big( G_{t0}, \; G_{t1}, \; ..., \;  G_{tn} \Big) \label{eq}
\end{equation}

where a Graph is a set of Nodes and a set of Edges.

\begin{equation}
G = \Big( N, \; E \Big) \label{eq}
\end{equation}

\subsection{GSTBN Model}
A geo-spatiotemporal network comprises a set of geo-spatiotemporal nodes and a set of geo-spatiotemporal edges. A geo-spatiotemporal node has a geographical longitude/latitude coordinate and may occur at select times or be persistent across all times. It may also move over time or remain stationary. A geo-spatiotemporal edge connects two nodes and may occur just once or multiple times or across all times. Geo-spatiotemporal edges have a numerical weight representing the geodesic distance between the linked nodes. This graph model assumes a bipartite network structure, ideal for mapping relations between two sets of nodes \cite{b9}.\\

\subsubsection{GSTBN Nodes} 
\hfill \break
Since this is a bipartite network, there are two types of nodes, observers and observables. Both types of nodes in this GSTBN represent geospatial coordinate points within the GoM, but they differ in the following ways, as outlined in this section.  \\

\paragraph{GCOOS Sensors (static)} 
\hfill \break
GCOOS sensor nodes are observer types. GCOOS sensor nodes are modeled as static or stationary, which means their geolocation is persistent across all time frames. The properties of each GCOOS node are in Table 1.  

\begin{table}[htbp]
\caption{\textbf{(GCOOS) Observer Node Properties}} 
\begin{center}

\begin{tabular}{c l} 
 \hline\hline
 \textbf{Label} & \textbf{Description} \\ [0.5ex] 
 \hline\hline
 id & Unique identifier number for each node \\ 
 \hline
 membership & Federal asset or local data node (ldn) asset \\
 \hline
 data source & Institution that operates the GCOOS sensor \\
 \hline
 platform & Name of observatory platform \\
 \hline
 mobility & Stationary or mobile \\  
 \hline
  geolocation & Latitude and Longitude of platform \\  
 \hline
 operational status & Active or Inactive  \\  
 \hline
 observations$^{\mathrm{a}}$ & Types of measures sampled by this platform  \\ 
 \hline
\end{tabular}
\end{center}
\footnotesize{$^{\mathrm{a}}$Observations were limited to those used by the HYCOM forecasting model, which are temperature, salinity, and ocean current velocities. }
\end{table}

\paragraph{HYCOM RoI Events (temporal)}
\hfill \break
HYCOM RoI nodes are observable types. HYCOM RoI nodes represent locations between consecutive snapshots where a significant change occurred in an observation. The properties for each RoI node are in Table II and Table III.  The residual formula quantifies the significance of the change between snapshots.   

\begin{table}[htbp]
\caption{\textbf{(HYCOM RoI) Event Node Properties }} 
\begin{center}

\begin{tabular}{c l} 
 \hline\hline
 \textbf{Label} & \textbf{Description} \\ [0.5ex] 
 \hline\hline
id & Unique identifier number for each node \\ 
 \hline
 geolocation & Latitude and Longitude of platform \\ 
 \hline
 RoI snapshots & Temporal dictionary with corresponding ROI data \\ & \textit{(see: Table III)}
 \\ 
 \hline
\end{tabular}
\end{center}
\end{table}

\begin{table}[htbp]
\caption{\textbf{ RoI Snapshot Properties }} 
\begin{center}

\begin{tabular}{c l} 
 \hline\hline
 \textbf{Label} & \textbf{Description} \\ [0.5ex] 
 \hline\hline
 snapshots & Nested Dictionary,
 outer keys are timestamps, \\ & inner keys are HYCOM observations.
\\ & The value is the residual score (real number) 
 \\ 
 \hline
\end{tabular}
\end{center}
\end{table}

\hspace{10mm} \textit{i) Residual formula} \\
In this paper, a residual is the squared difference between a given value from the same geospatial coordinate at two different times. Squaring the difference serves dual purposes. The first purpose is to ensure that the residual is always positive between the two times. The second purpose is to boost or diminish the residual based on the magnitude of its difference. If the difference is less than 1, it is diminished; if it is greater than 1, it is boosted. See Fig. 1 for a visualization of the residuals between two frames.  

\begin{equation}
residual = \big( value_{t_{n+1}} - value_{t_n} \big)^2
\end{equation}

\begin{figure}[h]
\centering
\includegraphics[width=0.4\textwidth]{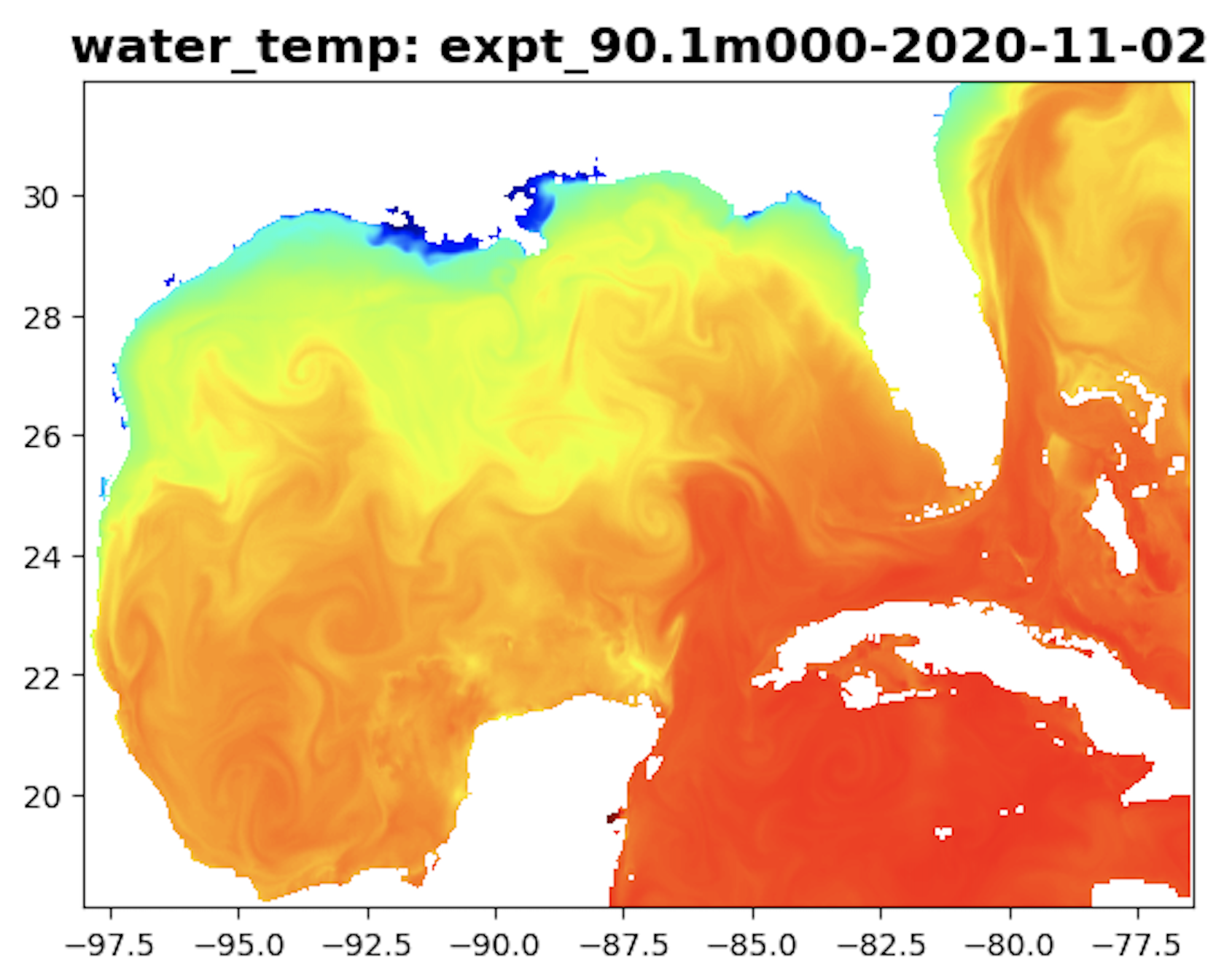}
\includegraphics[width=0.4\textwidth]{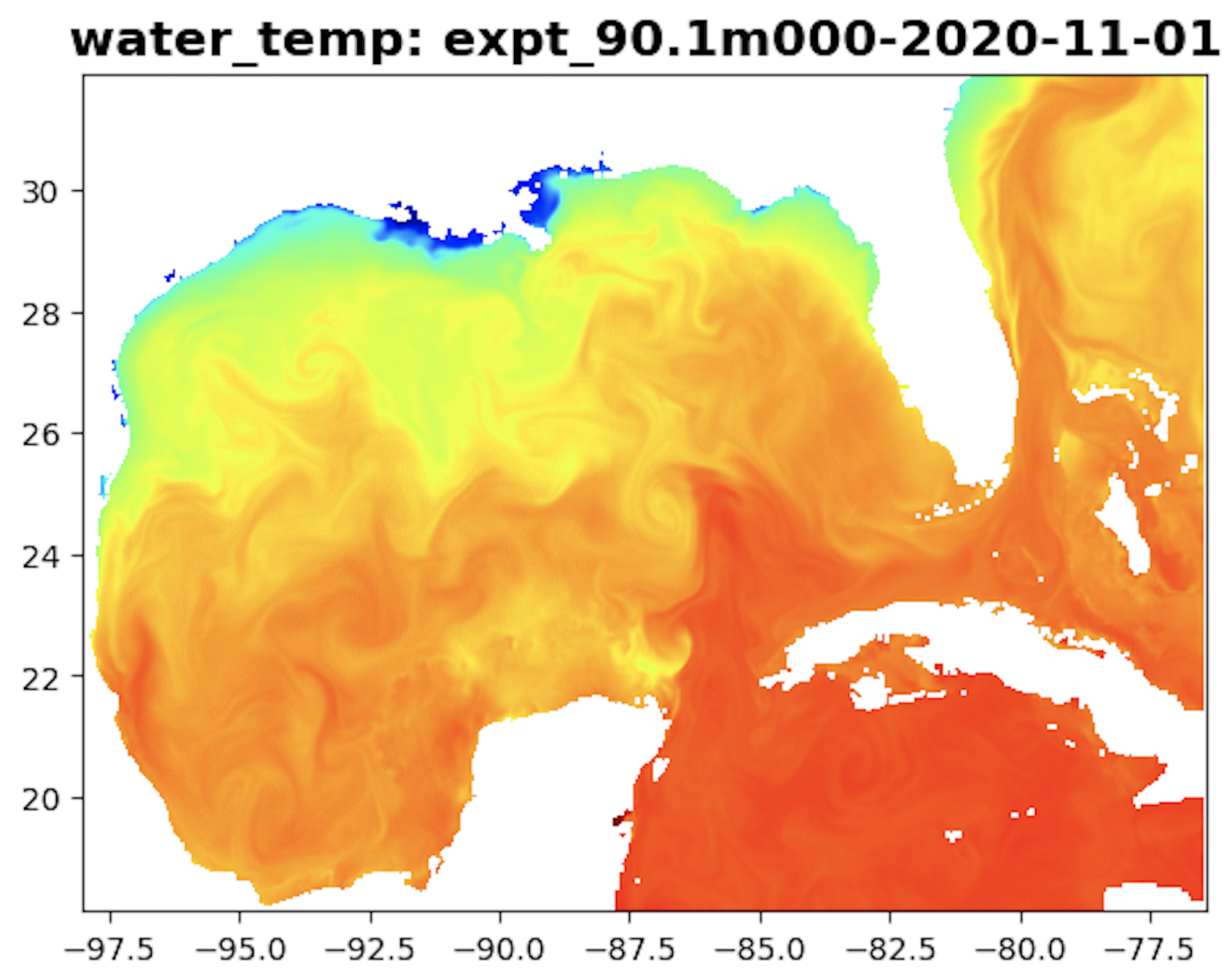}
\includegraphics[width=0.4\textwidth]{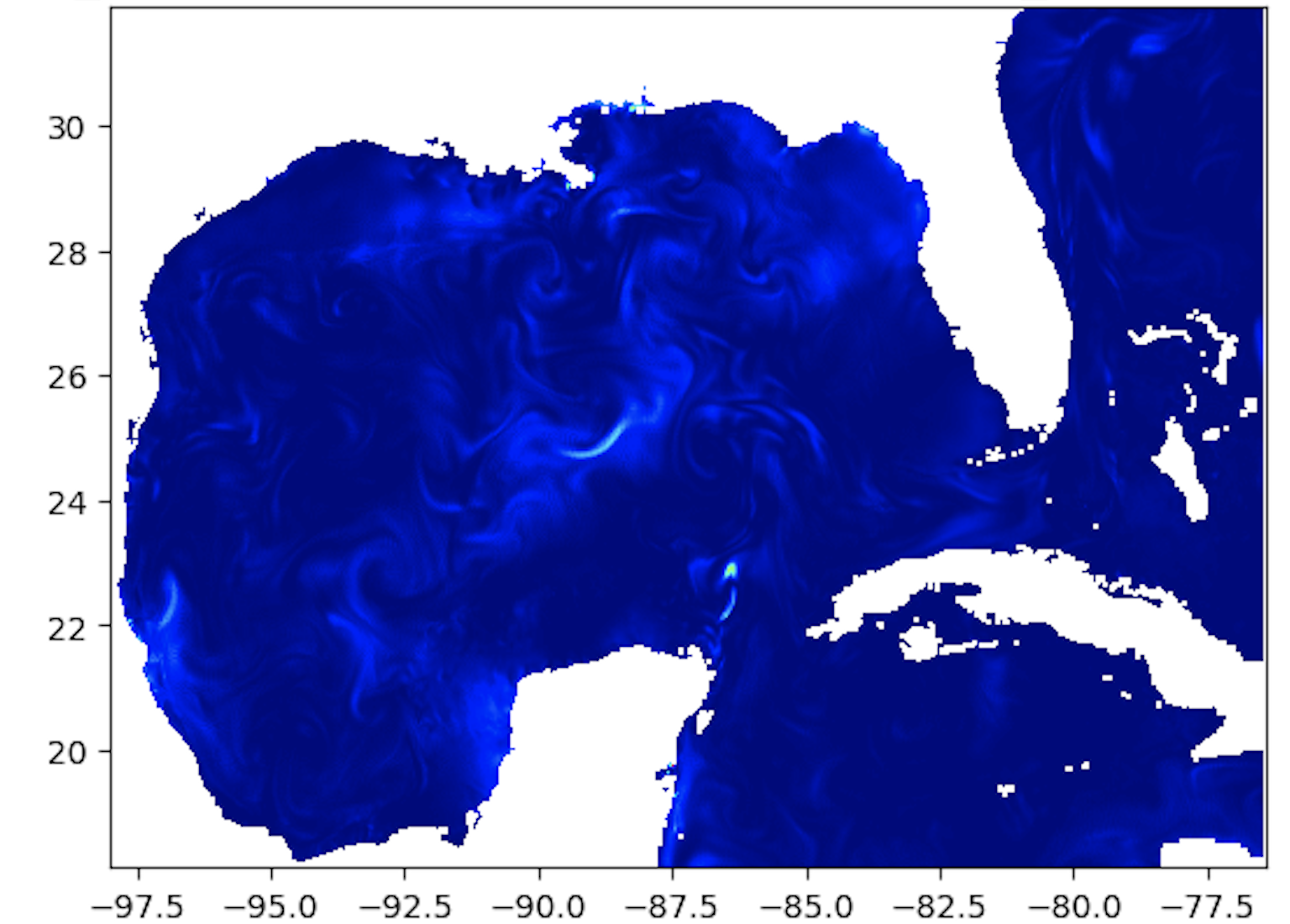}
\caption{Computing residuals between time frames}
\label{fig:Panel}
\end{figure}

In Fig. 1, the top and middle images are colormaps of the water temperature from the HYCOM model separated by a 24-hour period.  The bottom image is a colormap illustrating the residual difference between the two frames. These brighter locations within the bottom image depict the regions of interest for water temperature. \\

\hspace{10mm} \textit{ii) RoI formula} \\
The RoI is computed as the sum of residual values across all observations between time t\textsubscript{n} and t\textsubscript{n+1}. The resulting value is then compared to a thresholding value to determine if it is an RoI or ignored. The threshold value in Fig. 2 and Fig. 3 is 0.5.  

\begin{equation}
RoI= \sum_{v \in \text{ observations}} \left\{ \begin{array}{l} residual(v) \geq   \text{threshold} \\ \text{otherwise }  0 
\end{array}
\right\} 
\end{equation}
\\
\subsubsection{GSTBN Edges}
\hfill \break
GSTBN edges link GCOOS nodes and HYCOM nodes. Edge generation starts with an RoI Node and pairs with a GCOOS Node based on the shortest geodesic distance from the RoI node to the closest GCOOS node. The geodesic distance, in this case, is the spherical distance between two points, otherwise known as the "great circle distance" or “haversine” distance.

\begin{equation}\small{
\begin{array}{l}
dx = \sin\theta_{\text{LAT1}} * \sin\theta_{\text{LAT2}} \\
dy = \cos\theta_{\text{LAT1}} * \cos\theta_{\text{LAT2}} * \cos\theta_{\text{(LON1-LON2)}} \\
distance= \arccos(dx + dy) * R 
\end{array}
}
\end{equation}
Note:\\
\(distance\) = distance between two coordinates.\\
\(R\) = radius of Earth (approximately 6371.0090667KM)\\
\(\theta_{\text{LAT1}}\) = Latitude of the first coordinate in radians\\
\(\theta_{\text{LAT2}}\)= Latitude of the second coordinate in radians\\
\(\theta_{\text{LON1}}\)= Longitude of the first coordinate in radians\\
\(\theta_{\text{LON2}}\) = Longitude of the second coordinate in radians\\

\subsubsection{GSTBN Realizations}
\hfill \break
In Fig. 2, the GCOOS sensor nodes are colored red and remain stationary. In contrast, the HYCOM RoI nodes are grayscale with coloring relative to the strength of the residual and dynamic between frames.\\

\begin{figure}[h]
\centering
\includegraphics[width=0.4\textwidth]{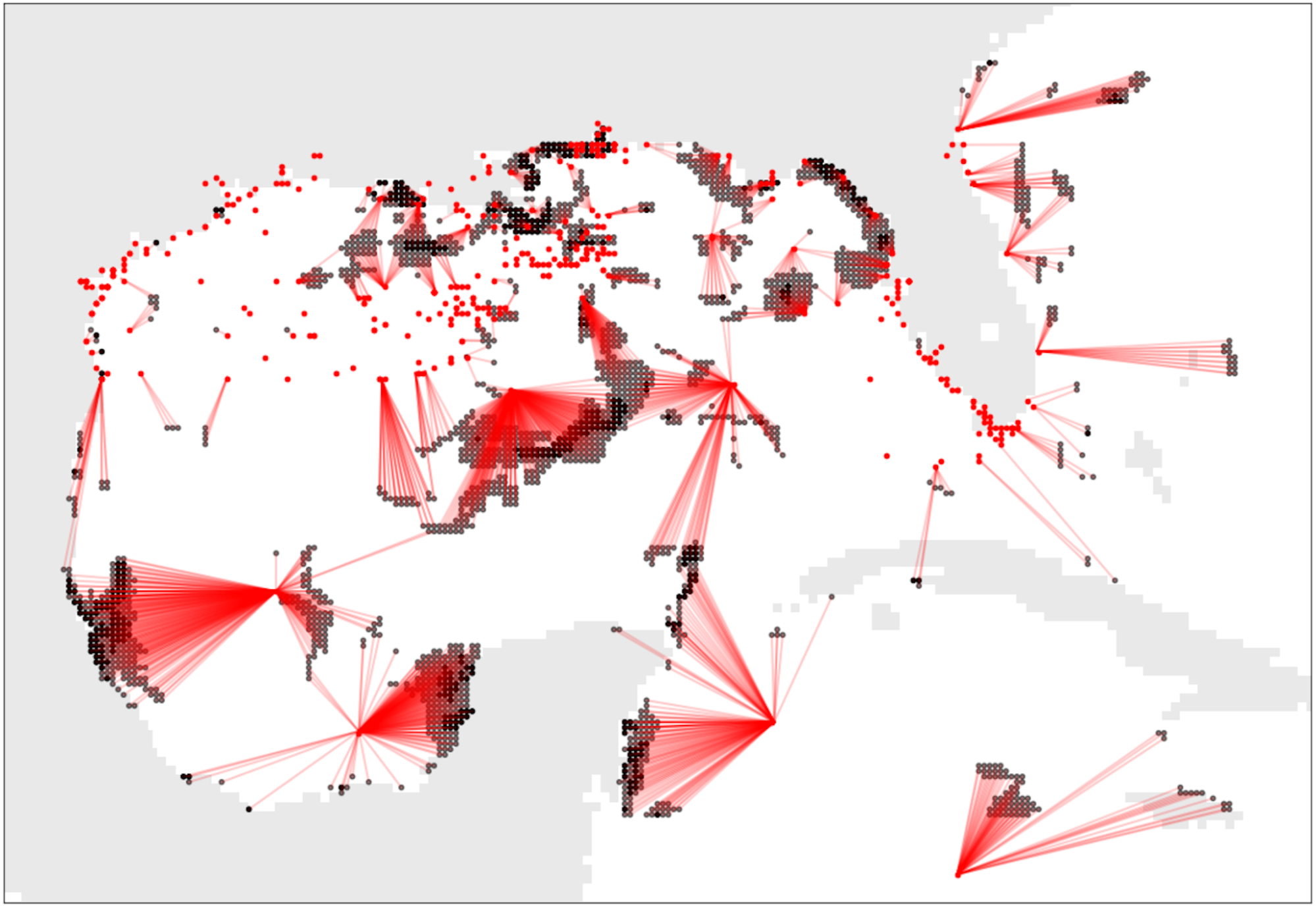}
\caption{ A snapshot of GSTBN at time \(t_0\). (Global view)}
\label{fig:Panel}
\end{figure}

In Fig. 3, the zoom window is the same position but each snapshot differs as the set of RoI nodes varies, illustrating the inherent stochastic problem of selecting optimal placements for new sensors.

\begin{figure}[h]
\centering
\includegraphics[width=0.4\textwidth]{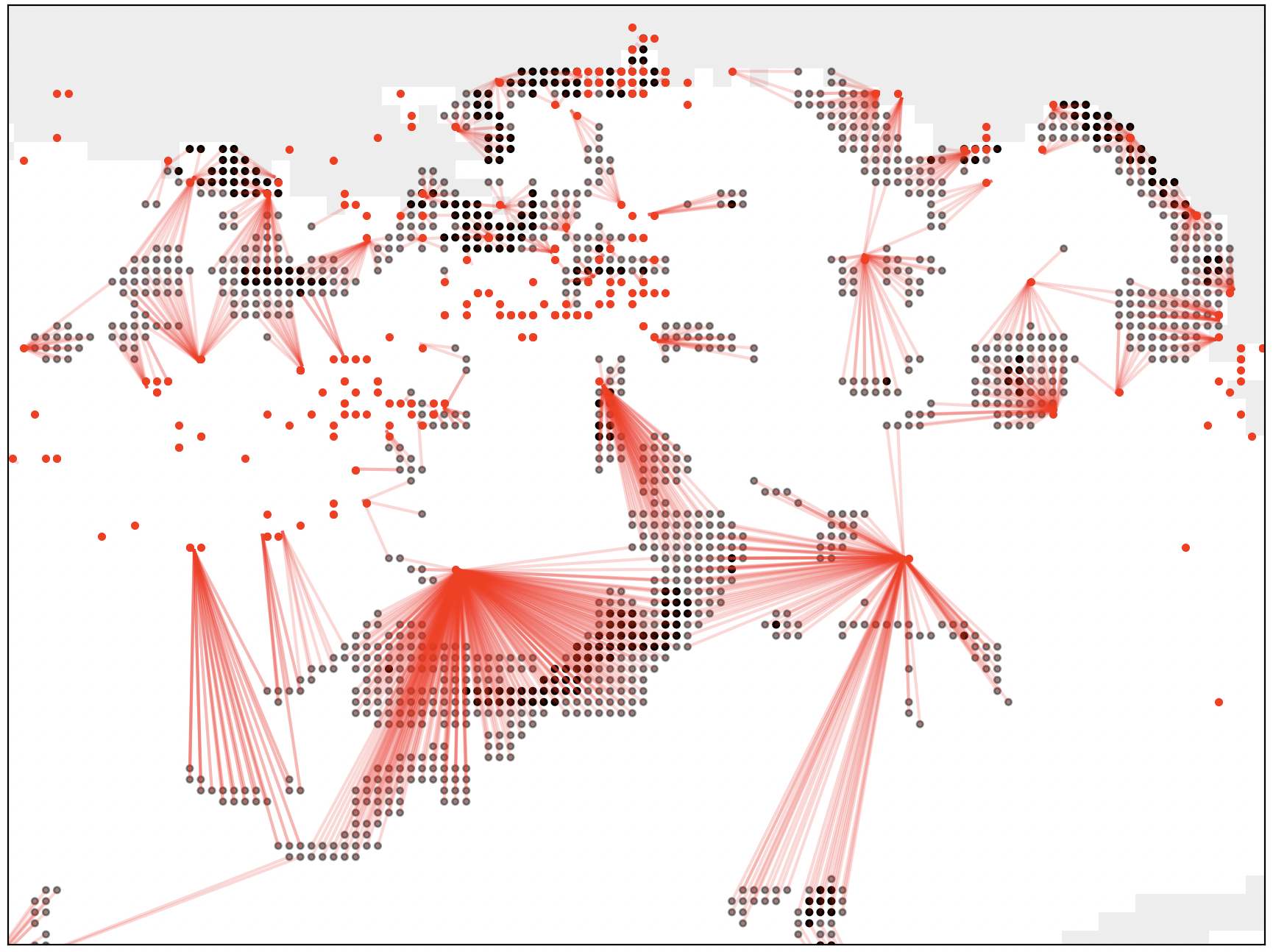}
\includegraphics[width=0.4\textwidth]{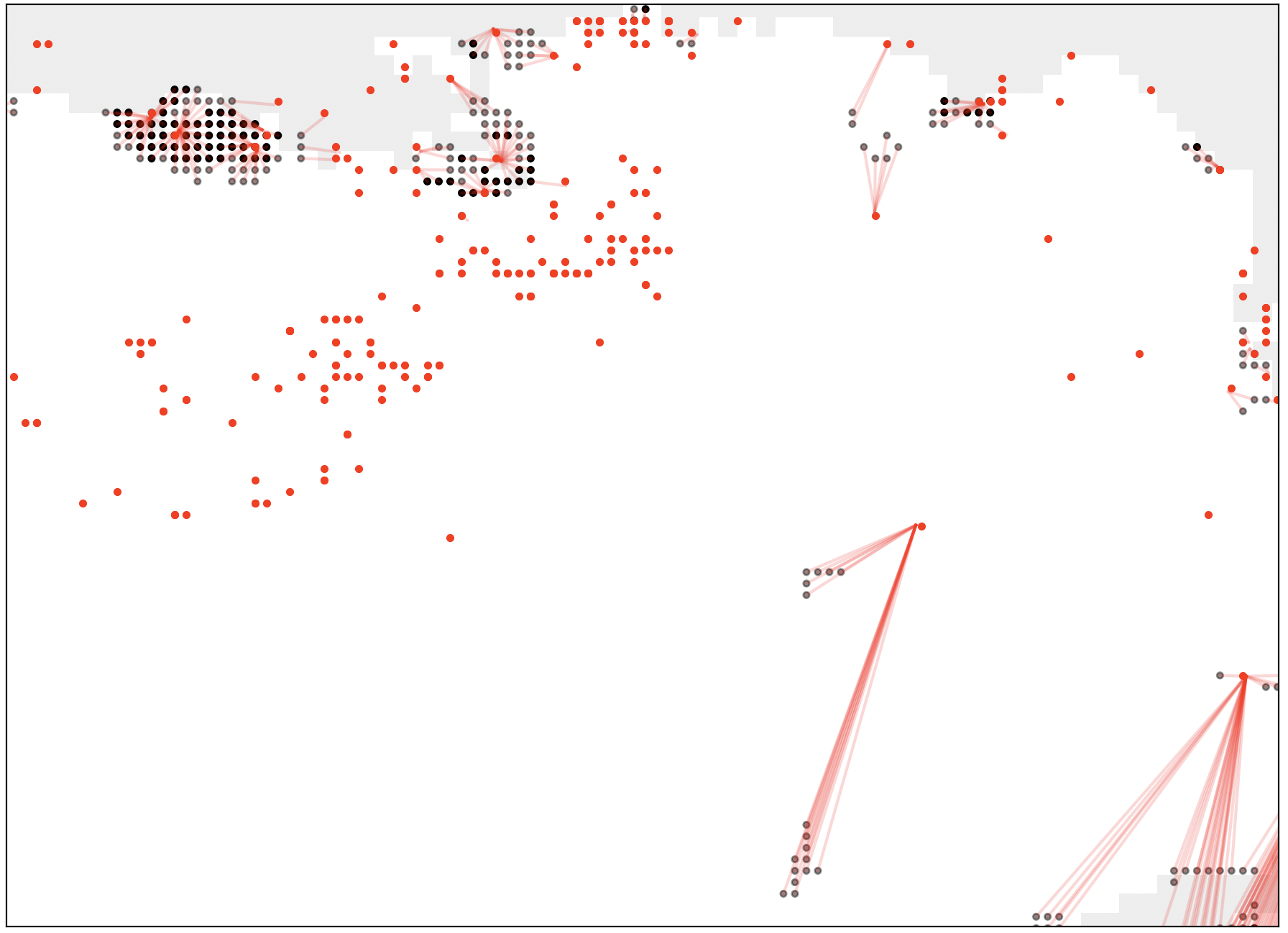}
\caption{Two temporal snapshots of the GSTBN at times \(t_0\) and \(t_2\) from same region. (Zoom view)}
\label{fig:Panel}
\end{figure}

\section{Analysis}
The goal of the GSTBN is to measure sensor placement strategies. Coverage and Coverage Robustness are the primary and secondary considerations used to measure the effectiveness of the spatial configuration of the GSTBN sensor nodes. These measures are defined in the subsections below. 

\subsection{Coverage measure}
This measure quantifies the coverage of the GCOOS sensor nodes to the HYCOM RoI nodes across all snapshots in the GSTBN. A maximal coverage would result in a GCOOS sensor node typically being close to an HYCOM RoI node. A suboptimal coverage could result in GCOOS sensor nodes being out of range from the HYCOM RoI nodes. \\

\subsubsection{Static Coverage }
\hfill \break
The static network coverage may be numerically computed as the sum of edge weights across all edges for a single discrete time step. This GSTBN encodes the geospatial distance between the closest GCOOS sensor node and an HYCOM RoI node as the edge weight. Given a set of weights, the formula below provides the coverage score for that timestep.  

\begin{equation}
coverage =  \sum edge \text{ } weights
\end{equation}

\vspace{2mm}

\subsubsection{Temporal Coverage }
\hfill \break
There are two approaches for measuring temporal network coverage by expanding the above definition for the static coverage measure: \\

\paragraph{Sum of Static Coverage Scores}
\hfill \break
The total temporal coverage is the sum of the static coverages across all timesteps within the GSTBN. 

\begin{equation}
total \text{ } temporal \text{ }  coverage = \sum_{t \in Timesteps} coverage(t)
\end{equation}

This approach may be unduly influenced by the presence of a bimodal distribution of coverage scores across the GSTBN timesteps.  One poor performance in a timestep heavily penalizes the total temporal coverage, or one positive performance greatly benefits it.   \\

\paragraph{Average of Static Coverage Scores}
\hfill \break
The average temporal coverage score better represents the expected coverage for any given time step within the GSTBN by using the quotient between the total temporal coverage and the total number of time steps. The Average Temporal Coverage is the primary measure used to rank the performance of the GSTBN configurations in this paper.

\begin{equation} \small {
average \text{ } temporal \text{ } coverage =  \frac {total \text{ } temporal \text{ } coverage} {number \text{ } of \text{ } timesteps} }
\end{equation}

\vspace{2mm}

\subsection{Coverage Robustness measure}
This measure quantifies the robustness of the GCOOS sensor nodes in its ability to cover the HYCOM RoI nodes adequately.  Maximal robustness results in a network configuration whereby the coverage would be minimally affected by removing a critical sensor node. With minimal robustness, removing a sensor node may substantially penalize the coverage score. Nodal centrality is a useful measure for determining the most critical sensor positions. To maximize robustness, the distribution of degree centrality should spread across multiple nodes instead of residing in only a few select critical nodes.  \\

\subsubsection{Static Degree Centrality }
\hfill \break
The static network degree centrality computes the distribution of edges across the GCOOS sensor nodes within the network. The distribution is the count of edges per node against the degree frequency across all nodes. \\

\subsubsection{Temporal Degree Centrality }
\hfill \break
There are two approaches to measuring the temporal network degree centrality.

\paragraph{Overall Centrality}
\hfill \break
The sum of all connections each node has through time over the entire temporal sequence \cite{b7}.

\paragraph{Per-Timestep Centrality}
\hfill \break
The sum of all connections each node has through time per time point
 \cite{b7}.   \\

Via  simulations, the network robustness is evaluated by removing the nodes with the highest nodal degree and recomputing the new network edges based on the revised distances to derive  the  new coverage score. If the coverage score increases significantly, the network is fragile to the loss of sensor nodes. If the coverage score remains relatively stable, the network is robust to the loss of sensor nodes.

\section{Optimizing Placements of New Sensor Nodes}
Identifying the optimal placements for new sensor nodes must start from the initial GCOOS sensor configuration. Given the stochastic nature of the HYCOM RoI nodes, a Monte Carlo simulation strategy determines the best locations for new GCOOS sensor nodes. The goal is to identify a nodal configuration that both distributes centrality and minimizes edge distances.

\subsection{Monte Carlo Simulation}
A Monte Carlo simulation is helpful to identify the probability of different outcomes in a non-deterministic environment due to the intervention of random variables \cite{b8}. In the case of this GSTBN, it is the HYCOM RoI node placements per timestep that are random. 

The Monte Carlo simulation begins by selecting a random coordinate within the spatial domain of the GSTBN. Then that coordinate is used to insert a new GCOOS sensor node into the GSTBN and recompute all of the edges. The updated edge list produces a new average temporal coverage score. That new coverage score compares against the current optimal placement's coverage score. If the new score is less than the current optimal score, then that random coordinate is saved at its optimal position. Repeating this process a suitably high number of times to exhaustively search the spatial space identifies the best location, which maximizes coverage across all timesteps in the testing dataset. 

An advantage of the Monte Carlo approach is that it is a distributed process at its core since each simulation is independent of the others. To identify optimal positions for multiple sensors within the GSTBN, perform this process sequentially, one node at a time. 

\section{Results}
All results use the Average Total Coverage measure. The score for the initial GCOOS sensor configuration is below. 

\subsection{Initial Sensors}
\begin{quote}
coverage score: 180222.806856 \\
\end{quote}

\begin{figure}[h]
\centering
\includegraphics[width=0.38\textwidth]{images/gstbn-init-t0.png}
\includegraphics[width=0.38\textwidth]{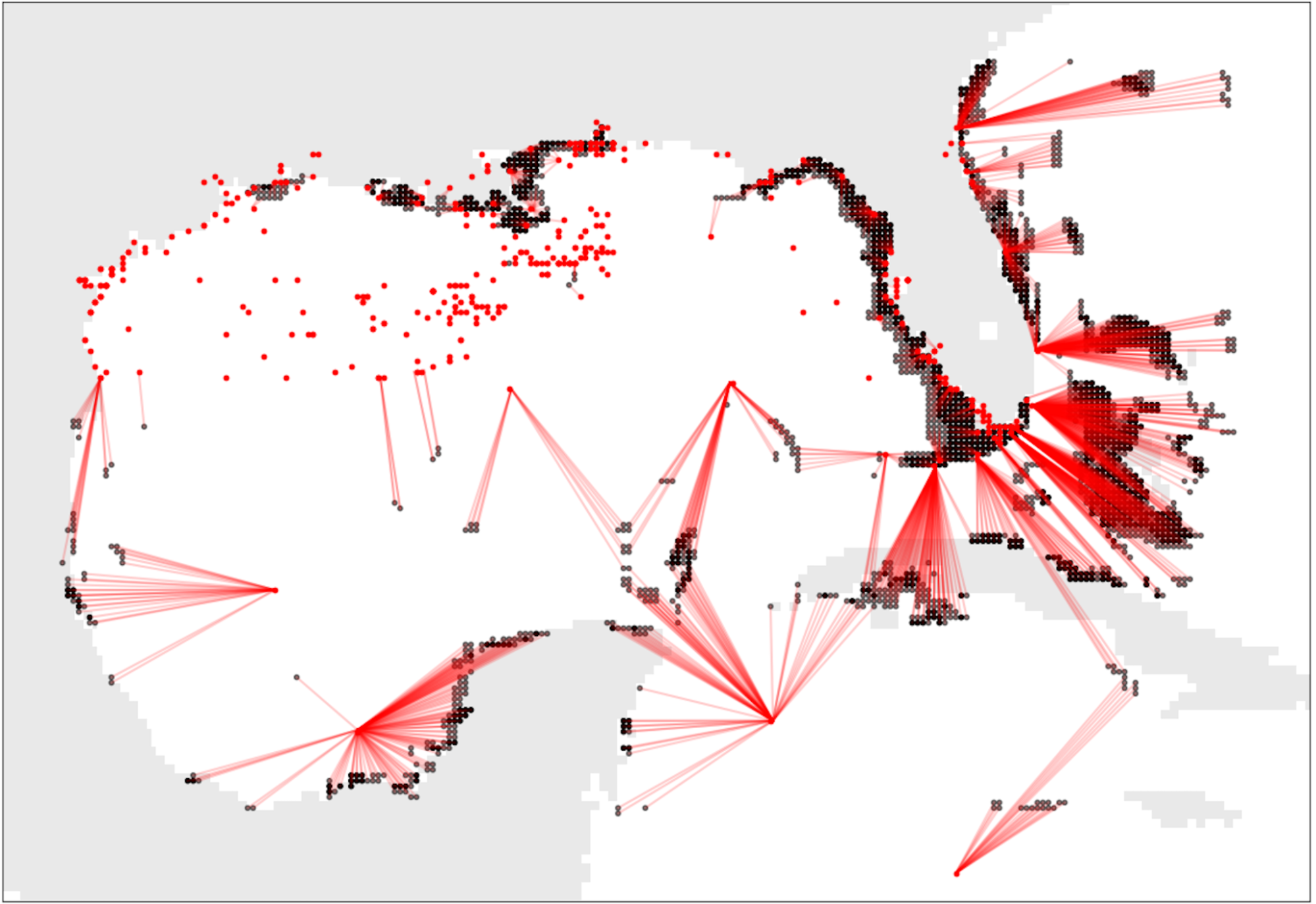}
\includegraphics[width=0.38\textwidth]{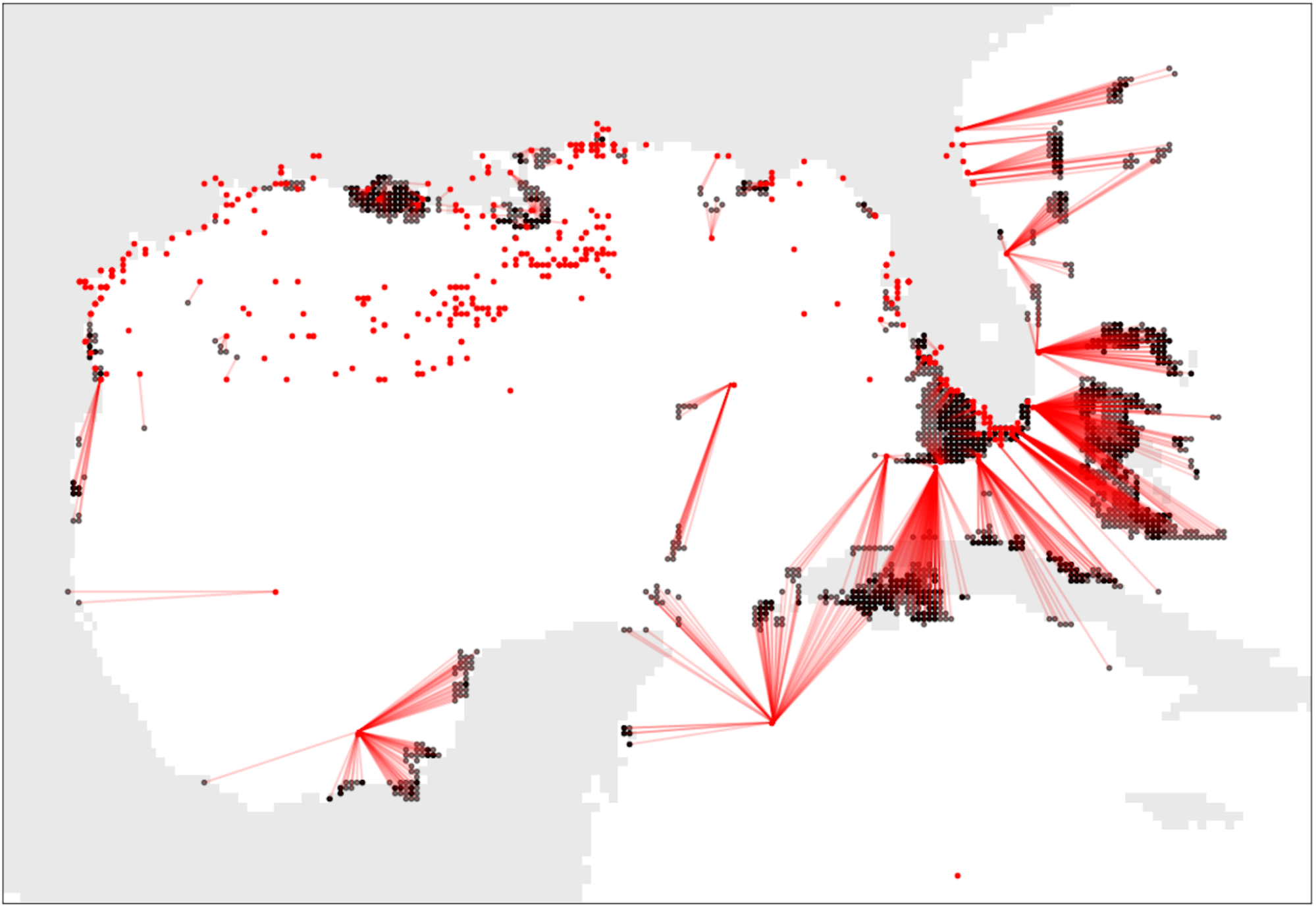}
\includegraphics[width=0.38\textwidth]{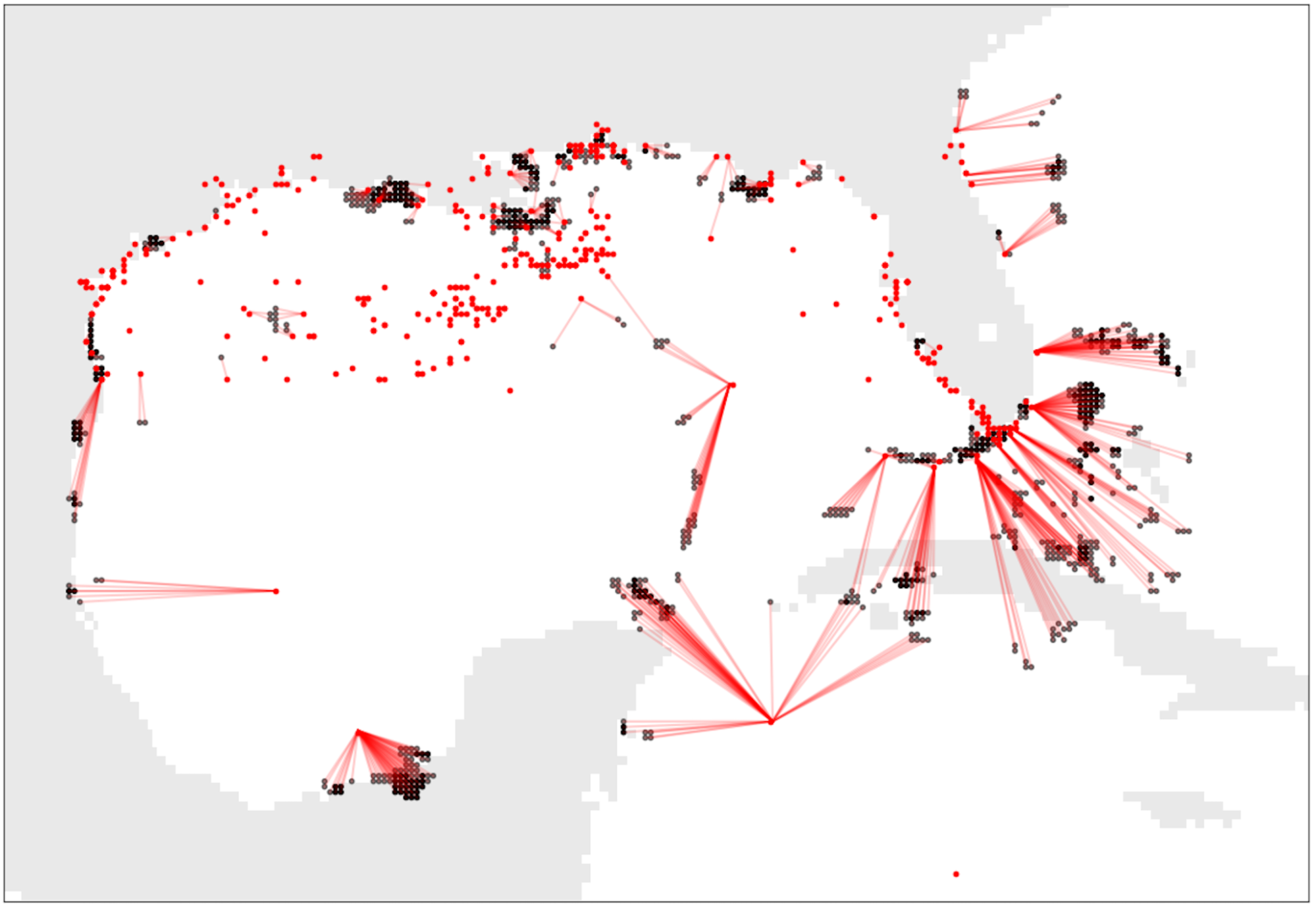}
\caption{Sequence of Temporal Snapshots of the GSTBN with the initial GCOOS sensor configuration}
\label{fig:Panel}
\end{figure}

\subsection{First New Sensor}
The Monte Carlo simulation  with 1,000 trials.
\begin{quote}
longitude: -78.7403109976445 \\ 
latitude: 24.385624429875215 \\
coverage score: 160873.88100 \\
\end{quote}

\begin{figure}[h]
\centering
\includegraphics[width=0.38\textwidth]{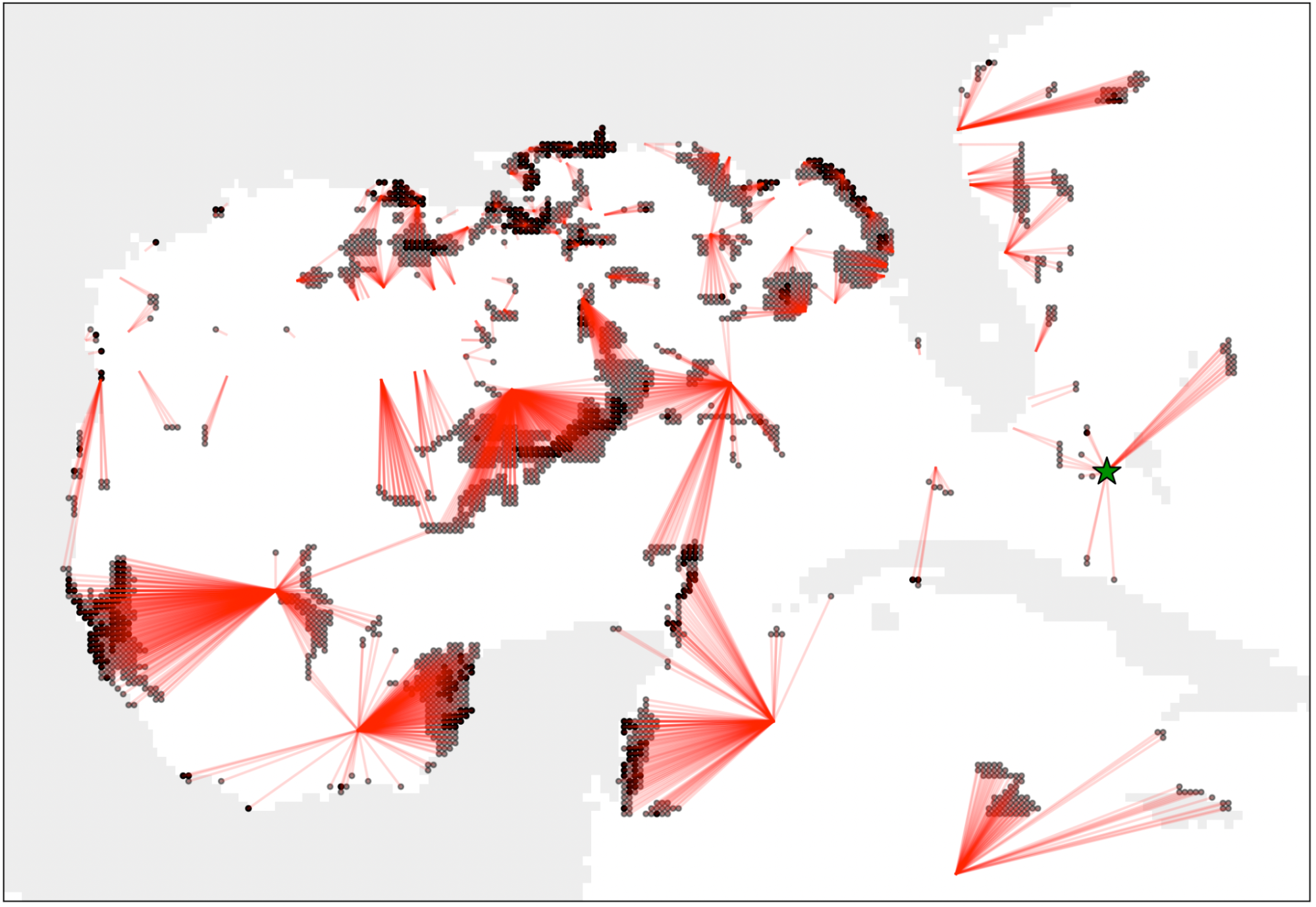}
\includegraphics[width=0.38\textwidth]{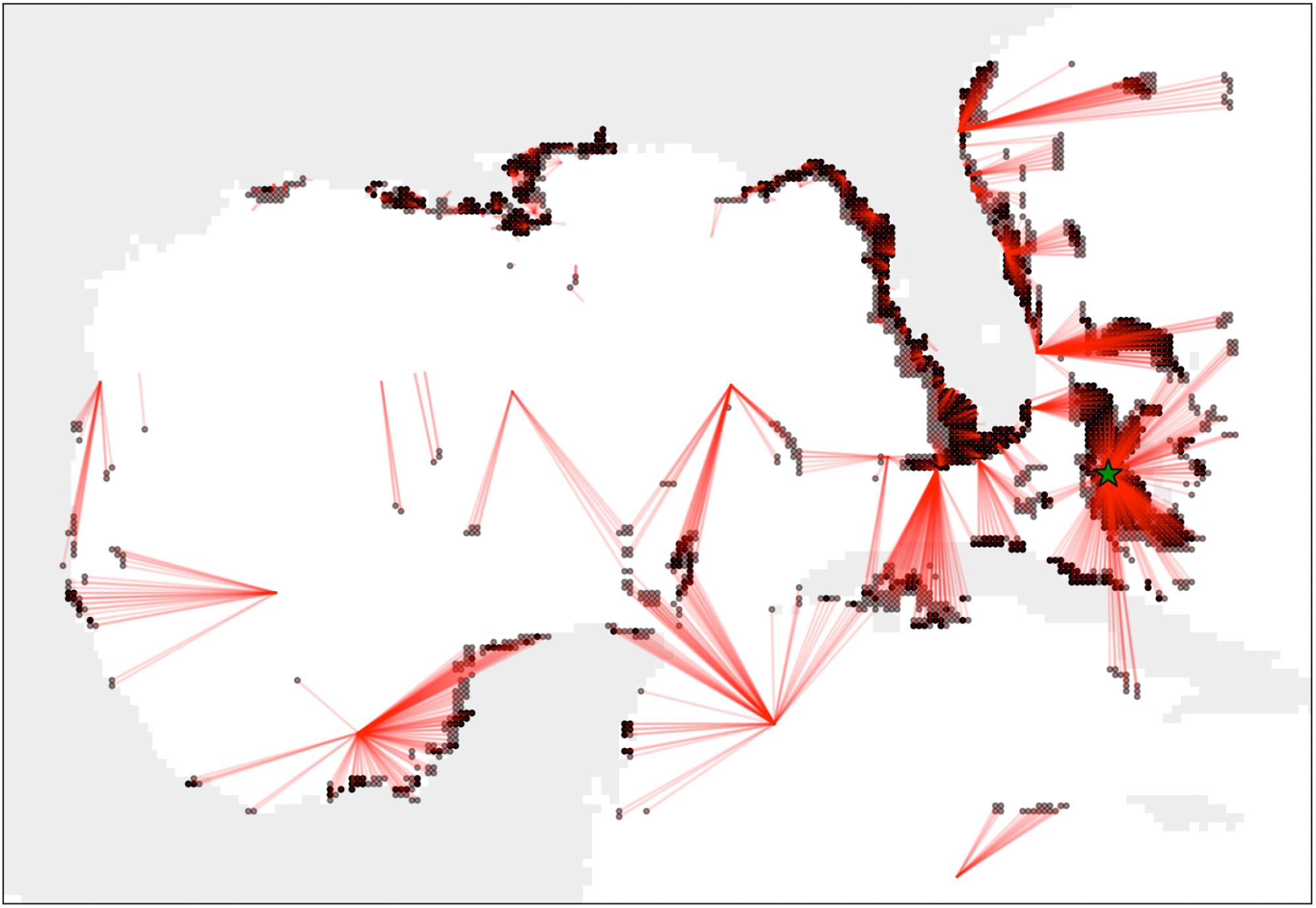}
\includegraphics[width=0.38\textwidth]{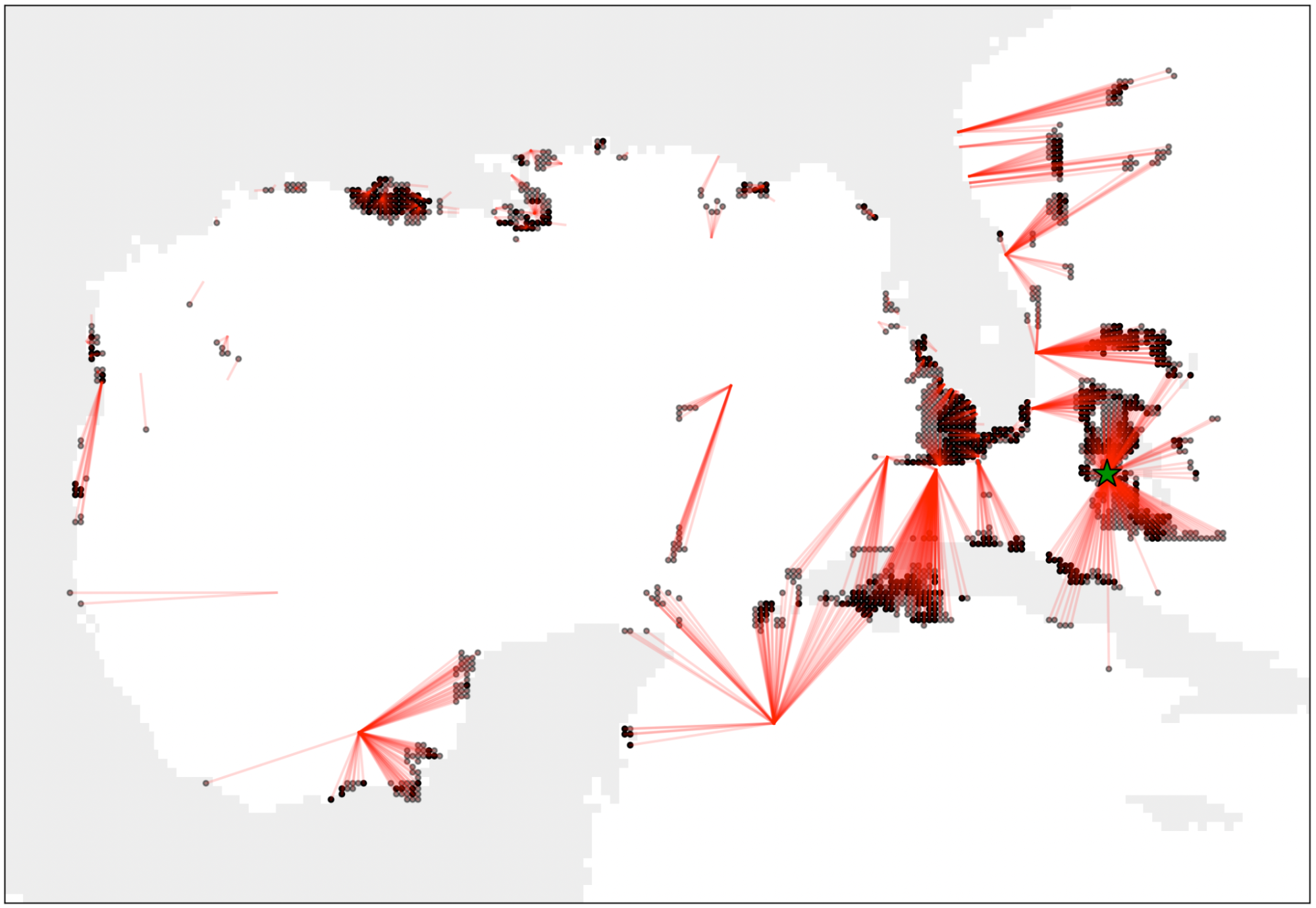}
\includegraphics[width=0.38\textwidth]{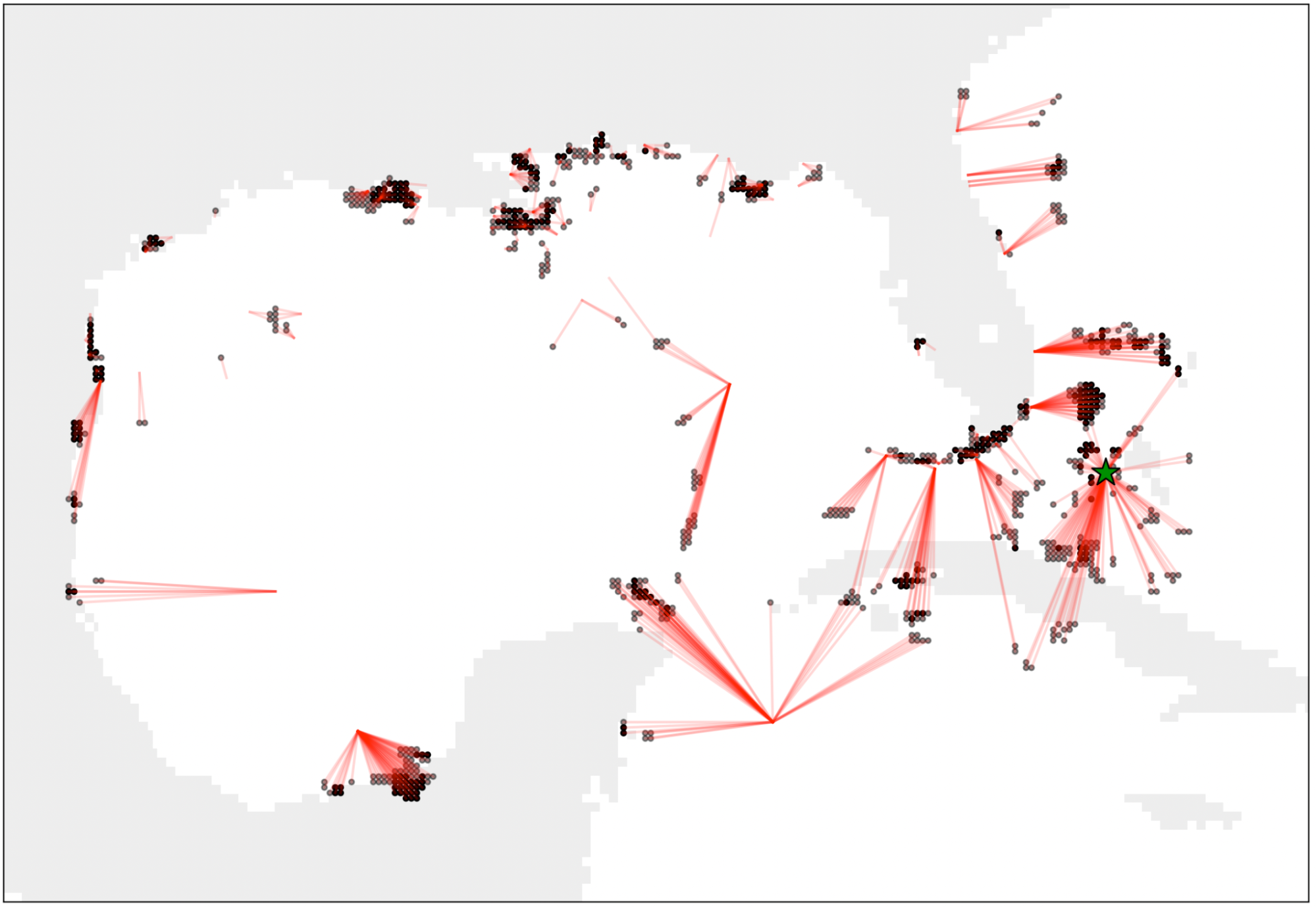}
\caption{Sequence of Temporal Snapshots of the GSTBN with the suggested position for a new sensor represented as a green star}
\label{fig:Panel}
\end{figure}

\subsection{Second New Sensor}
The Monte Carlo simulation  with 1,000 trials. 

\begin{quote}
longitude: -85.81532374804107 \\
latitude: 22.561994782989117 \\
coverage score: 147411.742470\\
\end{quote}

\begin{figure}[h]
\centering
\includegraphics[width=0.38\textwidth]{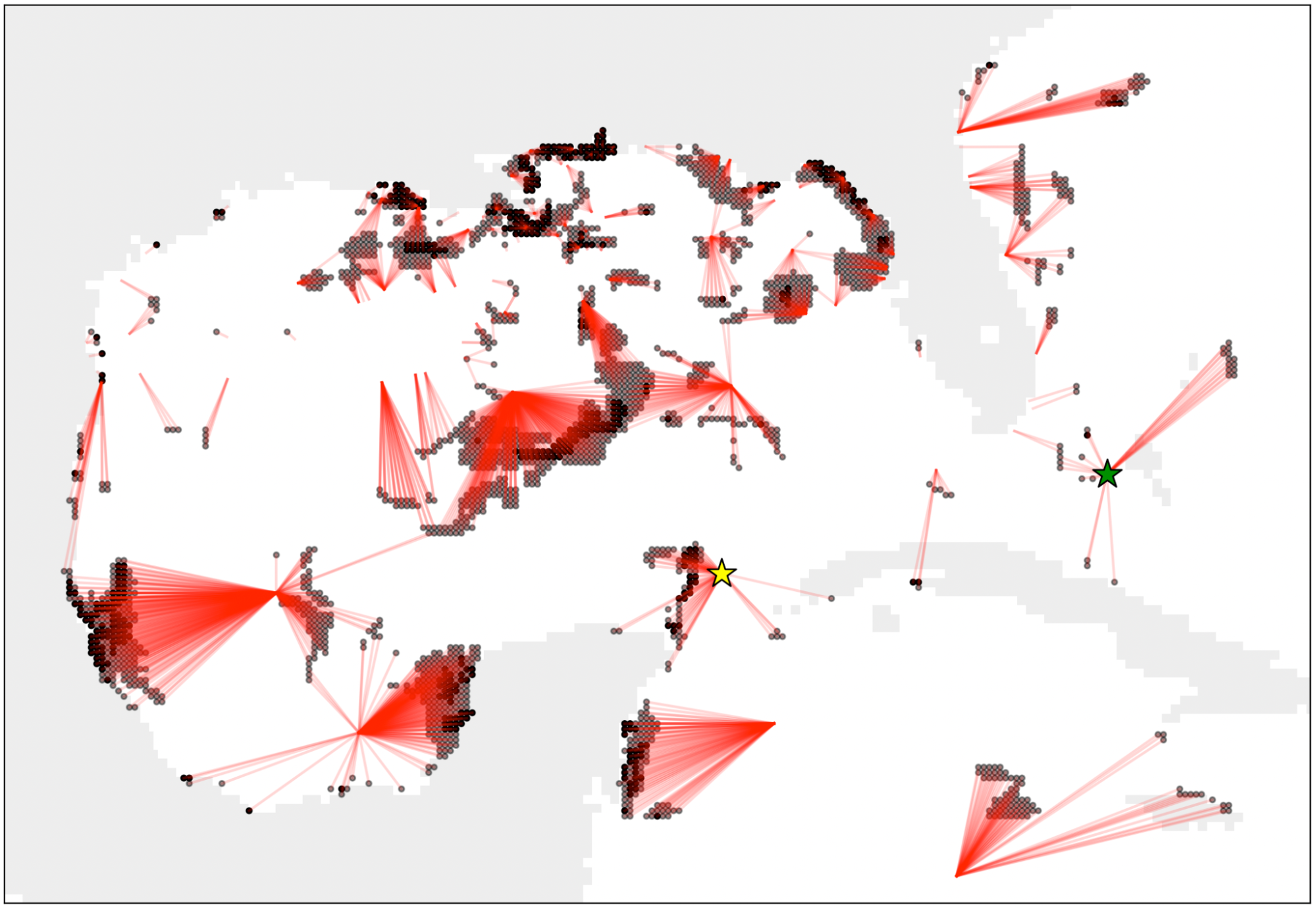}
\includegraphics[width=0.38\textwidth]{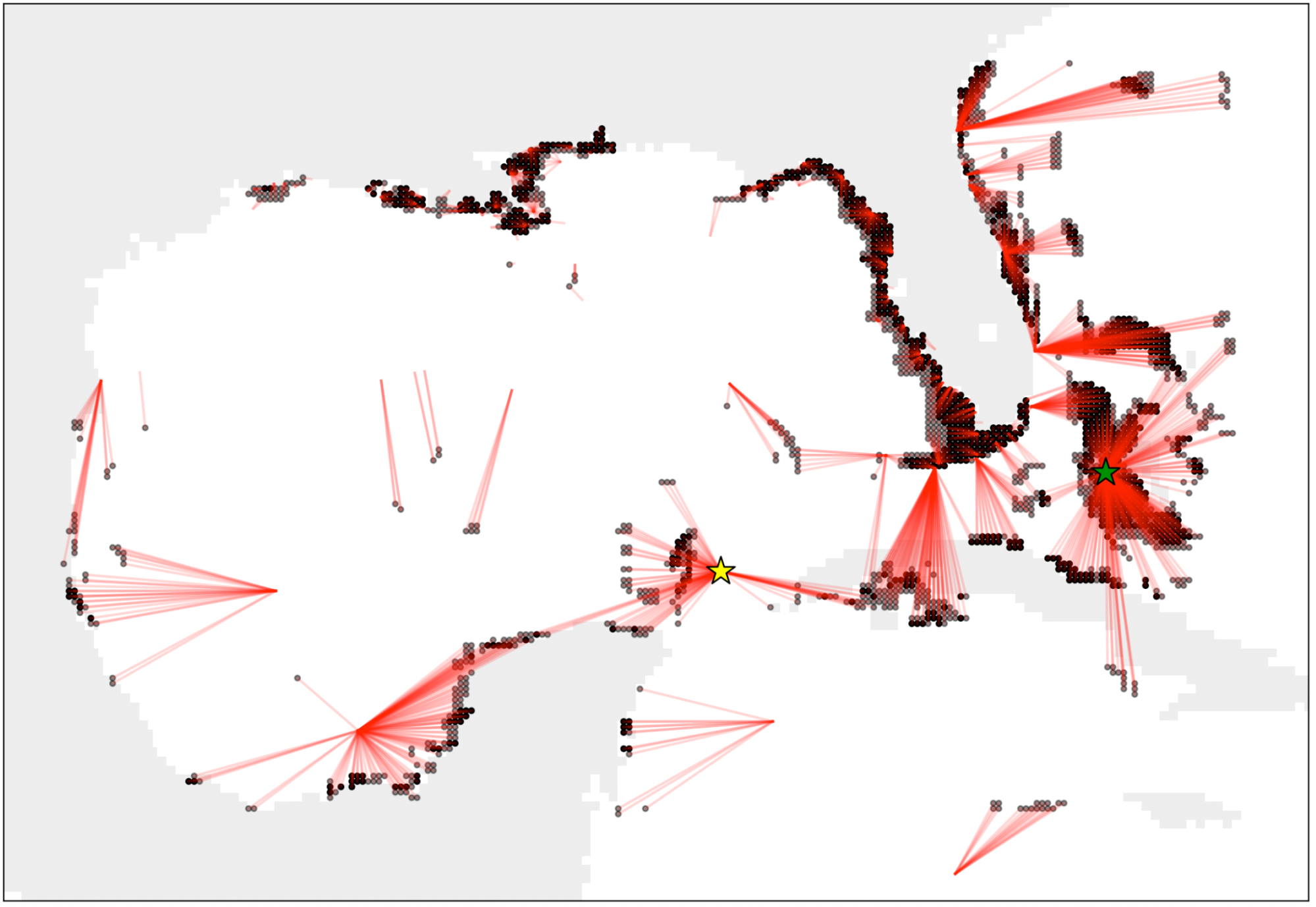}
\includegraphics[width=0.38\textwidth]{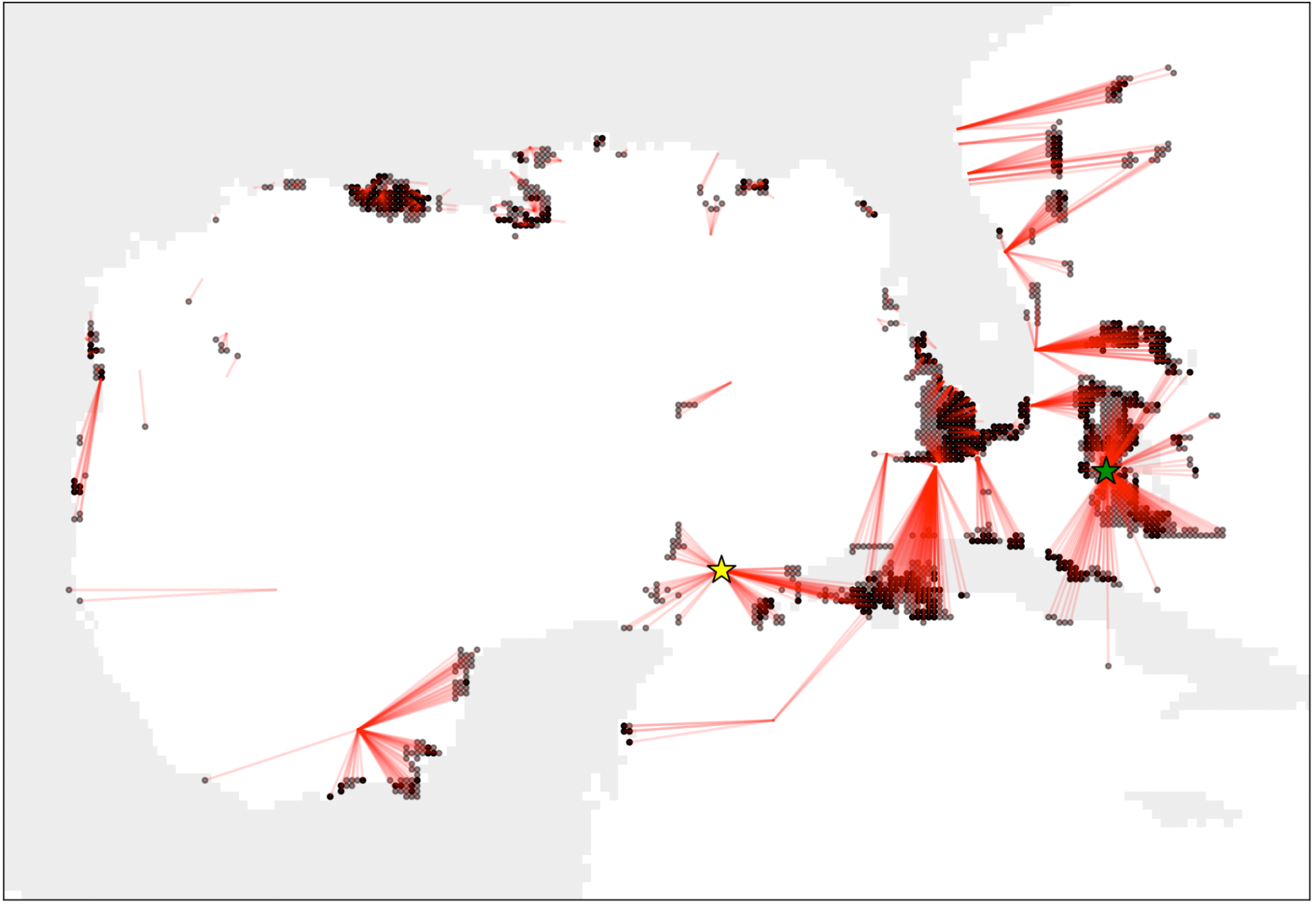}
\includegraphics[width=0.38\textwidth]{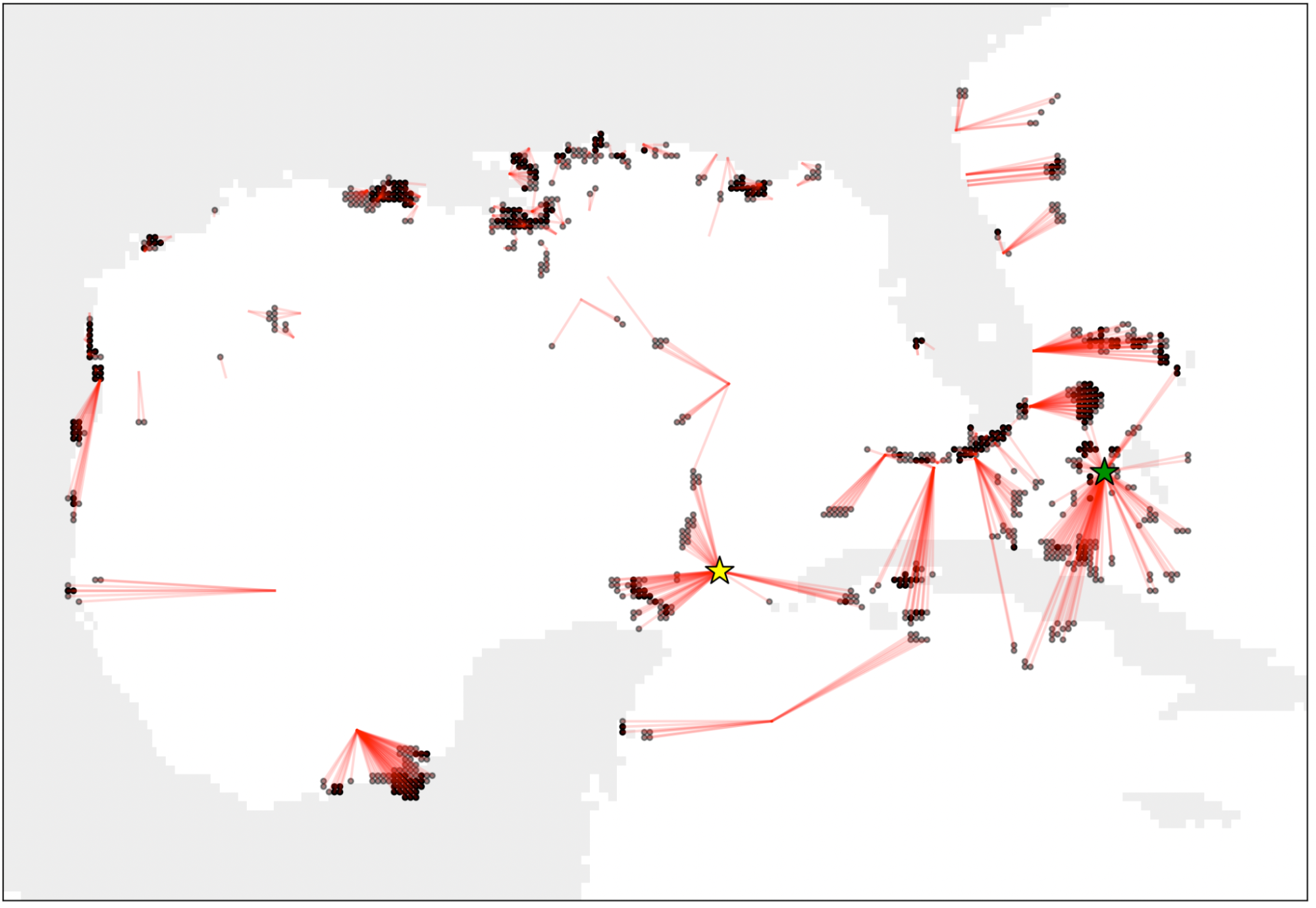}
\caption{Sequence of Temporal Snapshots of the GSTBN with the suggested position for a 2nd sensor represented as a yellow star}
\label{fig:Panel}
\end{figure}
\clearpage

\section{Discussion}
The results illustrate how this approach uses a bipartite network model to capture the coverage performance between persistent observers and observable stochastic events within a spatial environment and successfully identify near-optimal locations to expand the observer array. 

As seen in the Results section, the coverage score significantly decreases with the insertion of both the  first and second new nodes. Notice that the placement of each new node maximizes its coverage across all times.  

The results identified through this process would then support any management decisions from the platform/station operators/owners when they intend to add new observational equipment. 

We can expand the guidance provided with even more information by performing community clustering on the observer or the observable nodes. For example, cluster the GCOOS platform nodes into different communities based on their platform ownership status. Cluster the HYCOM RoI nodes together in communities based on their observation type. The GSTBN can model such constraints; however, these levels of fine detail and consideration are omitted from the scope of this paper.

\section*{Acknowledgments}
This work was partly supported by the U.S. Department of the Navy, Office of Naval Research (ONR), and Naval Research Laboratory (NRL) under contracts N00173-20-2-C007 and N00173-20-2-C007, respectively.\\


\end{document}